\documentclass[12pt]{article}
\begin{document}
\begin{center}
{\large\bf
Periodicity and quark--antiquark static potential}
\vglue .2in
{Pong Youl Pac}
\vglue .2in
{\it  Department of Physics, Seoul National University\\
Seoul, 151-742,  Korea}\\
{\it  E-mail : parkso@nuri.net}\\
\end{center}         
\vspace{2cm}
\begin{abstract}
\indent 
Beyond the standard model, a static potential between quark pairs is obtained
 phenomenologically (QCD inspired), associated with the range of strong 
interaction, when the virtual exchange gluon squared momentum transfer has 
a periodicity for periodic boundary conditions of the quark--pair system 
enclosed by a constant volume,  
in the lowest order of the
{\it effective} perturbed QCD (in which the gluon  propagator is  replaced 
by  the {\it  effective}  gluon one).  This  potential includes a 
periodicity dependent effect, characterized by a finite
face value of the periodicity $N$, in addition to the periodicity
independent potential (the Coulomb type plus linear one).
That periodicity dependent effect, dominant at short distance,
is applied to an explanation of the top quark mass
$$m_t=8\pi m_\pi N^{\frac{1}{2}},$$
whose numerically calculated results indicate approximately both upper and lower bounds 
of $m_t$
$$177~\mbox{\mbox{GeV}}> m_t > 173 ~\mbox{\mbox{GeV}}$$
for the range of strong interaction $L=1.40~fm~(=m_\pi^{-1})$.\\

\end{abstract}

\newpage
\noindent{\large \bf 1. Introduction}\\


The range of strong interaction $L$ is phenomenologically considered 
as a constant $m^{-1}_\pi $, the Compton wave length of the pion, 
which is consistent with the experimental proton charge radius \cite{1} :
$$ L=m^{-1}_\pi = \mbox{a constant}.$$                       
\noindent 
This may  be a standard length  
of the quark pair system, which  is available for determining 
the quark sizes, i.e., the quark masses (the Compton wave length).\\
\indent 
In this system, a particle (quark pair) with its momentum
$k_n$ (specified by $n$) is confined in the range $r_n$, where this $r_n$
satisfies the relation
\begin{equation}
r_nk_n=\frac{1}{2}~~~~ (\mbox{the position momentum uncertainty})\label{1}
\end{equation}
for the static case. We are interested in a quark pair enclosed by
a cubic volume $L^3$, with periodic boundary conditions for momenta
 [$k'_i=\frac{2\pi}{L}\eta_i~~~ (i=1,2,3 ; \eta_i=0,\pm1,\pm2,\cdots$)].
The above momentum points are finite, equal spacing and periodic for 
periodic boundary conditions of the quark pair system  \cite{2}.
This description  may be  extended formally to  the one  of the 
squared momentum transfer in the Minkowski momentum space and hence 
the effective gluon propagator may be derived in the fashion,
as seen in section 2 of this work. Then, the static potential
between quark pairs in the quantum view may be obtained from the analyticity
of the effective gluon propagator for periodic boundary conditions of
the quark pair system, as seen in the succeeding sections. This static potential 
includes a periodicity dependent effect, characterized 
by a finite face value of the periodicity $N$,
in addition to the periodicity independent potential 
(the Coulomb type plus linear one), where the periodicity 
dependency of potential is defined by  the  periodicity 
condition for the  static case, as seen  in  sections 3 and 4. 
That periodicity dependent effect, dominant at short distance, 
may be used to determine the heavy quark masses.\\
\indent Recently, the  top quark of about 180 \mbox{GeV}  was discovered
at Fermi lab experimentally \cite{3}.
At present, it may be important for us to find a way in explaining 
the top quark mass for a further theory beyond the standard model 
\cite{4},\cite{5}. All known phenomenological
potentials for quark pairs (purely phenomenological or QCD inspired) 
are the flavor independent ones, but by these potentials no one has
been successful in explaining the top quark mass \cite{6},\cite{7}. 
The existence of the (heavy) top quark may suggest that there exists 
another kind of static potential between quark pairs which includes,
e.g., a periodicity dependent effect, characterized 
by the finite face value of the periodicity, in addition to the 
periodicity independent potential (the Coulomb type plus linear one).\\
\indent The aim  of this work is to build  up a phenomenological model
(QCD inspired) beyond the standard model which enables us to derive 
a static potential between 
quark pairs, including the periodicity dependent effect 
(for explaining the top quark mass) in addition to the periodicity 
independent potential (the Coulomb type plus linear one).
Our attention is concentrated on the result, $L=8\pi 
RN^{\frac12}$ (or $m_t=8\pi m_\pi N^{\frac12}$) which is the  multiple of
the top  quark confined range $R$  and the square root  of the top 
quark confined range number (the finite face value of the periodicity) 
$N^\frac12$ excepting $8\pi$. 
This $N$ is determined from the periodicity dependent effect
for $R$. Since $L~(=m^{-1} _\pi)$ is a constant, $R~(=m^{-1} _t)$ is 
obtained from $L=8\pi RN^{1\over2}$ if $N$ is determined as mentioned above.
Our above discussion is true for the $l$- quark pair, e.g., 
$L= 8 \pi R_l N_l ^{\frac12} $, where $l=u(d)$, $s, c, b$ and $t$,
respectively. This model is developed within the framework of the 
quark model with three generations.\\
\indent This paper is organized as follows : In section 2, 
an effective gluon propagator is derived as an average of the gluon
propagator in the perturbed QCD on the finite and equal spacing (arbitrary) 
squared momentum transfers (periodicity recurrent points intervals) for 
periodic boundary conditions of the quark pair system.
In section 3, the periodicity independent potential 
(the Coulomb type plus linear one) is obtained from the above fashion
on the finite and equal spacing (arbitrary) squared momentum 
transfers (periodicity recurrent points intervals) under the periodicity 
condition for the static case. 
In section 4, the static potential between quark pairs, 
as done in the previous section, is derived when the gluon squared momentum
transfers (periodicity recurrent points intervals) are finite, equal spacing (definite) and periodic for  
periodic boundary conditions of the quark pair system.
This static potential includes the periodicity dependent
effect in addition to the periodicity independent
one. That periodicity dependent effect produces a relation    
$m_t=8\pi m_\pi N^{\frac12}$ (or $L=8\pi RN^{\frac12}$).
Section 5 is devoted to the explanation of the top quark mass 
$m_t$, approximately $177 ~\mbox{GeV}> m_t > 173 ~\mbox{GeV}$, for the range of strong
interaction $L=1.40~ f_m(=m^{-1}_\pi)$ by the periodicity 
dependent effect for $R$. Section 6 is left for the concluding remarks. 
In Appendix, the excitation probability in section 2 is derived.
Throughout this paper, the particle size is expressed by the Compton
wave length.\\
\vspace{0.5cm}

\noindent {\large \bf 2. Effective Gluon Propagator}\\

\indent  A static  potential between a quark pair is a function of $r$,
where $r$ is the distance between the center of the quark and the 
one of its antiquark, with the Compton wave length radius $\frac12r_n$ 
($r_n \equiv r_{\bar n}$ (the Compton wave length of anti--particle)),
respectively. The closest distance between the quark pair is defined as
\begin{equation}
r=\frac{1}{2} r_n+\frac{1}{2} r_n=r_n,\label{2}
\end{equation}
which enables us to obtain the corresponding quark mass $m_n$ 
(the Compton wave length) and assures there is no static potential 
inside this $r_n$ (the closest distance). This means the Compton 
wave length of a quark (fermion) equals to the closest distance between 
the quark--pair (boson). \\
\indent According to the field theory \cite{2}, if the quark pair 
at the closest distance, which behaves like one bosonic particle, 
is enclosed by the cubic volume $L^3$, the plane wave functions 
of this system,
$\exp(i\vec{k'}\cdot\vec{r'})=\exp(i\vec{k}\cdot\vec{r})$
if $k'_i=\frac{1}{2} k_i,  ~r'_i=2r_i$, form a complete set and
the periodic conditions show that $k'_i=\frac{2\pi}{L} \eta_i 
=\frac{1}{2}k_i$. The choice of $r_i=\frac{1}{2}r'_i$
(and $k_i=2k'_i$) may be adequate to the lattice point in the 3-dimensional
Euclidean space  if $ r'_1=r'_2=r'_3=L$. The  above discussion may be  
extended formally to the one in the 4-dimensional Euclidean space 
which is connected to the Minkowski momentum space, $k_4'=-ik_0'$ (real)\cite{8}. \\
\indent Introducing such a lattice point in the 4-dimensional Euclidean space
$$n'\equiv{\eta'}^2=\eta^2_1+\eta^2_2+\eta^2_3+\eta^2_4,$$
where $\eta_4=0,\pm1,\pm2,\cdots$ and $n' =0,1,2,\cdots$ (by definition),
respectively, the squared momentum transfer in the Minkowski momentum 
space is 
\begin{eqnarray}                     
               -k^2\equiv k^2_E\!&=&k^2_1+k^2_2+k^2_3+k^2_4
                            =\left({\frac{2\pi}{\frac{1}{2}L}}\right)^2\!
                            (\eta^2_1+\eta^2_2+\eta^2_3+\eta^2_4)\nonumber\\
                         &=&c^2n'\left(\frac{cL}{4\pi}\right)^{-2}\!
                           =c^2n'a_c^{-2}\ge 0~(a_c\equiv cL{(4\pi)}^{-1}),\label{3}
\end{eqnarray}
where $k_4=(\frac{2\pi}{\frac{1}{2}L})\eta_4$ (similarly to 
$k_i=(\frac{2\pi}{\frac{1}{2}L})\eta_i)$,
$c$ is a constant (without loss of generality $1\le c <\infty$) 
and the subscript $E$ means Euclidean, respectively. From Eq.(\ref{3}),
we have the following periodicity condition 
\begin{equation}
-a^2_c k^2=c^2n'~~~~~~~ (n'=0,1,2,\cdots),\label{4}
\end{equation}
for $-k^2$. For the static case, we have the following 
periodicity condition for $|\vec k|^2$
\setcounter{equation}{3}
\renewcommand{\theequation}{\arabic{equation}$'$}
\begin{equation}
a^2_c{|\vec{k}|}^2=c^2n ~~~~~~~ (n=0,1,2,\cdots), \label{4'}
\end{equation}
where ${|\vec{k}|}^2=k^2_1+k^2_2+k^2_3$ and
      $n\equiv\eta^2=\eta^2_1+\eta^2_2+\eta^2_3 (n'=n+\eta^2_4.n'=n,~\mbox{if}~ \eta_4=0.)$,
respectively. We call  this $n$ as the periodicity and the largest
$n=N$ as the finite face value of the periodicity, vice versa. 

\indent Let us start with the gluon propagator in the perturbed QCD \\
\setcounter{equation}{4}
\renewcommand{\theequation}{\arabic{equation}}
\begin{equation}
-iD^{\alpha\beta}_{\mu\nu}(k)=-i\left(\frac{\delta_{\alpha\beta}}{k^2}\right)
 \left[ g_{\mu\nu}-(1-\xi)\frac{k_{\mu}k_\nu}{k^2}\right].\label{5}
\end{equation}
\noindent For $\xi=1$, instead of Eq.(\ref{5}), we may start with\\
\setcounter{equation}{4}
\renewcommand{\theequation}{\arabic{equation}$'$}
\begin{equation}
-iD^{\alpha\beta}_{\mu\nu} (k)=-i\frac{\delta_{\alpha\beta}g_{\mu\nu}}{k^2}.\label{5'}
\end{equation}
\noindent In Eq.(\ref{5}) or Eq.(\ref{5'}), $k^2$ in all denominators
should be considered as $k^2+i\epsilon~~(\epsilon \to 0^{+})$, 
because of the causal property of the gluon propagator.
When the hadronic system (characterized in terms of both the squared 
momentum transfer and the  finite length parameter $a_c$) is introduced,
the intervals between the squared momentum transfer periodicity recurrent 
points are finite and equal spacing (arbitrary) 
as seen in Eq.(\ref{4}). The gluon propagator (keeping $\delta_{\alpha\beta}
g_{\mu\nu}$ out) $-k^{-2}$ is the inverse of the 
``squared momentum transfer operator $-k^{2}$''
in this gauge \cite{8}. This means that both the finite and equal 
spacing squared momentum transfer and the corresponding gluon 
propagator have the same excitation probability between the respective
finite and equal spacing levels. It is to be stressed that the periodicity 
of the gluon squared momentum transfer may be represented by the one of 
$-k^2$.\\
\indent The excitation probability between the gluon propagators 
$(n'+1)(-k^{-2})$ and $(n'+2)(-k^{-2})~~~(n'=0,1,2,\cdots)$ 
is $\exp(a^2_ck^2)$ as shown in Appendix.
This $\exp(a^2_ck^2) ~(1>\exp(a^2_ck^2)>0 ~~\mbox{for}~~ 0\!<\!-k^2\!<\!\infty)$~
is the most probable one according to the information theory \cite{9}.
Thus for the finite and equal spacing periodic squared momentum transfer 
recurrent points intervals case, this model requires 
the average $(-k^{-2})$ as
\setcounter{equation}{5}
\renewcommand{\theequation}{\arabic{equation}}
\begin{eqnarray}
(-k^{-2})_{av}
   &=&\left(1\cdot(-k^{-2})+2\cdot(-k^{-2})
              \exp(a^2_ck^2)+\cdots \right.\nonumber\\
   & &\left.+(n'+1)(-k^{-2})\exp(n'a^2_ck^2)+\cdots \right)
      \left(\sum_{n'=0}^\infty\exp(n'a^2_ck^2) 
      \right)^{-1}\nonumber\\
   &=&(-k^{-2})\sum_{n'=0}^\infty(n'+1)\exp(n'a^2_ck^2) 
      \left(\sum_{n'=0}^\infty\exp(n'a^2_ck^2)
      \right)^{-1}\nonumber\\
   &=&(-k^{-2})\left\{X\frac{\partial}{\partial X}\sum_{n'=0}^\infty
      X^{n'}+\sum_{n'=0}^\infty X^{n'}
               \right\}
      \left(\sum_{n'=0}^\infty X^{n'}
      \right)^{-1}\nonumber\\
   &=&(-k^{-2}) (1-X)^{-1}\label{6}
\end{eqnarray}
instead of $(-k^{-2})$, owing to the periodicity condition for $(-k^2)$, Eq.(\ref{4}),
where $X=\exp(a^2_c k^2)$.
Noting the original form of ${D^{\alpha\beta}_{\mu\nu}}(k)$, 
we have the following effective gluon propagator
\begin{equation}
-iD^{\alpha\beta}_{\mu\nu}(a_c,k)=\left(-i\delta_{\alpha\beta}k^{-2}
         \left(1-\exp(a^2_c k^2)\right)^{-1} \right)\left[ g_{\mu\nu}
            -(1-\xi)k_\mu k_\nu k^{-2}\right],\label{7}
\end{equation}
which behaves like the gluon propagator of the perturbed (the $k^{-4}$ 
type \cite{10}) QCD in the infinite (the zero) squared momentum transfer limit.
If we put 
\begin{equation}
f(-a^2_c k^2)\equiv 1-\exp(a^2_c k^2),\label{8}
\end{equation}
this $f(-a^2_c k^2)$ may be an important weight function describing 
the hadronic structure.
In fact, since $\lim\limits_{-k^2\to\infty}  f(-a^2_c  k^2)=1$  and
$\lim\limits_{-k^2\to  0} f(-a^2_c k^2)= \lim\limits_{-k^2\to 0}(-a^2_c k^2)$,
we have $\lim\limits_{-k^2\to\infty}\left(-k^{-2} f^{-1}(-a^2_c k^2)\right)
        = \lim\limits_{-k^2\to\infty}(-k^{-2})$ and 
         $\lim\limits_{-k^2 \to 0}\left(-k^{-2}f^{-1}(-a^2_c k^2)\right)
        =\lim\limits_{-k^2\to 0}(a^{-2}_c k^{-4})$.\\ 

Finally, we may get
\begin{equation}
-iD^{\alpha\beta}_{\mu\nu}(a_c,k)=\left(-i\delta_{\alpha\beta}k^{-2}f^{-1}
   (-a^2_c k^2)\right)\left[ g_{\mu\nu}-(1-\xi)k_\mu k_\nu k^{-2}\right].\label{9}
\end{equation}  
For $n'\!=\!N'\!\gg\!1$, Eq.(\ref{8}) may be still applicable (approximately)
to our calculation, since such $n'$ is considered as continuous.\\       

\vspace{.5cm}
\noindent {\large\bf 3. Periodicity Independent Static Potential}\\

\indent To obtain the periodicity independent static potential between 
quark pairs, we have to use the effective gluon propagator Eq.(\ref{7}),
because of the periodicity condition, Eq.(\ref{4'}). For the primitive
form of static quark-antiquark potential in our model (based on the 
quantum views), we can start with the integral 
\begin{equation}
I=-(2\pi)^{-3}\!\int_0^\infty\!|\vec{k}|^2 d|\vec{k}|\!\int_{-1}^1\! d(\cos\theta)
   \!\int_0^{2\pi}\! d\phi\exp(i\vec{k}\cdot\vec{r})|\vec{k}|^{-2}\!
   \left(1\!-\!\exp(-a^2_c|\vec{k}|^2)\right)^{-1}\!,\label{10}
\end{equation}
where we have used Eq.(\ref{7}) with $\xi=1$ and have kept 
$\delta_{\alpha\beta}g_{\mu\nu}$ out. This $I$ may be reduced to the usual QCD 
case if $a_c \to \infty$, i.e., $L \to \infty$.
After elementary calculations, including the exchange $|\vec{k}|\to-|\vec{k}|$,
we have                                              
\begin{equation}
I=-(2\pi r)^{-1}(2\pi i)^{-1}\int_{-\infty}^\infty 
   d|\vec{k}|\exp(i|\vec{k}|r)|\vec{k}|^{-1}
    \left(1-\exp(-a^2_c|\vec{k}|^2)\right)^{-1}.\label{11}
\end{equation}
This integral, which is analytic in the upper half plane of complex 
momentum (according to the Jordan's lemma \cite{11}), is rewritten as
\begin{equation}
I=-(2\pi r)^{-1}(2\pi i)^{-1}\oint_\Gamma dz \exp(izr)z^{-1}
   \left(1-\exp(-a^2_c z^2)\right)^{-1},\label{12}
\end{equation}
where $z\equiv |\vec{k}|\exp(i\psi)(0\!\le\!\psi\!\le\!\pi)$.
In Eq.(\ref{12}), a triple pole at $z=0$ (in Eq.(\ref{4'}), $n=0$, i.e., 
the {\it periodicity independent} case) is enclosed within the contour
integral in the upper half plane of complex momentum. 
Some residue calculations give us 
\begin{equation}
I\!=\!-(2\pi r)^{-1}(2\pi i)^{-1}\!\oint_\Gamma dz             
\exp(izr)z^{-1}\left(1\!-\!\exp(-a^2_c z^2)\right)^{-1}\equiv\!
\left(-(2\pi r)^{-1}\right)a_{-1}, \label{13}
\end{equation}
where
\begin{equation}
a_{-1}=\frac12(1-a^{-2}_c r^2)+(2r)(\pi a^2_c)^{-1}
(\lambda^{-1})_{\lambda \to 0}.\label{14}
\end{equation}
Thus we have
\begin{equation}
I=(4\pi)^{-1}(-r^{-1}+a^{-2}_c r)-(\pi^2 a^2_c)^{-1}
(\lambda^{-1})_{\lambda\to 0}.\label{15}
\end{equation}
In Eq.(\ref{15}), the last term is infrared divergent ($\lambda=0$),
but is independent of r. So we will neglect this term throughout
our discussion. After a normalization, $\frac{4}{3}g^2_s$,
we find the following periodicity independent potential between
quark pairs, in the lowest order 
\begin{equation}
V(a_c,r)=\frac{4}{3}g^2_s I=\alpha_s(-r^{-1}+a^{-2}_c r),\label{16}
\end{equation}
where $\alpha_s=\displaystyle\frac{4}{3}(\frac{g^2_s}{4\pi})$ and $a_c$ 
is defined as in Eq.(\ref{3}). This $V(a_c,r)$ is just the periodicity 
independent potential, which is similar to the Coulomb type plus 
linear one \cite{6}.\\ 
\indent If we take a pure phenomenological weight function
\begin{equation}
f_p(-a^2_c k^2)= -a^2_c k^2 (1-a^2_c k^2)^{-1} \label{17}
\end{equation}
instead of $f(-a^2_c k^2)$ in Eq.(\ref{8}), then we have
\begin{equation}
-k^{-2}f_p^{-1}(-a^2_c k^2)= -k^{-2}+a^{-2}_c k^{-4} \label{18}
\end{equation}
and from Eq.(\ref{18}) we obtain the static quark-antiquark potential 
$V_p(a_c,r)=\alpha_s(-r^{-1} + a^{-2}_c r)$ in the lowest order.
Our weight function, Eq.(\ref{8}) is rewritten as
\begin{equation}
f(-a^2_c k^2)=f_p(-a^2_c k^2)+(1-a^2_c k^2)^{-1}
\left\{1-g(-a^2_c k^2)\right\}, \label{19}
\end{equation}
where $g(-a^2_c k^2)=(1-a^2_c k^2)\exp(a^2_c k^2),
~ \lim\limits_{-k^2\to 0}g(-a^2_c k^2)=1$ and
$\lim\limits_{-k^2\to\infty}g(-a^2_c k^2)= 0$,
respectively. Thus, it is found that
$\lim\limits_{-k^2\to 0}f(-a^2_c k^2)
=\!\lim\limits_{-k^2\to 0}f_p(-a^2_c k^2)\!=\! 0$ and 
$\lim\limits_{-k^2\to \infty}f(-a^2_c k^2)
=\!\lim\limits_{-k^2\to\infty}f_p(-a^2_c k^2)\!=\!1$, respectively.\\

\vspace{.5cm}
\noindent {\large\bf 4. Static Potential with Periodicity Dependent Effect}\\

The periodicity condition, Eq.(\ref{4}) or Eq.(\ref{4'}) shows 
that the intervals between the gluon squared
momentum transfer periodicity recurrent points are finite and equal 
spacing (arbitrary). If the constant $c$ is chosen as
\begin{equation}
c=(2\pi)^\frac12, \label{20}
\end{equation}
then we have the periodic condition
\begin{equation}
-a^2 k^2 = 2\pi n'~~~~~
(n' = 0,1,2,\cdots ~:~a=2^{-1}(2\pi)^{-\frac{1}{2}}L), \label{21}
\end{equation}
which means that the intervals between points of $-k^2=\displaystyle\frac{2\pi}{a^2}n'$ are finite, equal spacing 
and periodic for periodic boundary conditions of the quark pair system. For the static case,
we have the following periodic condition 
\setcounter{equation}{20}
\renewcommand{\theequation}{\arabic{equation}$'$}
\begin{equation}
a^2|\vec{k}|^2=2\pi n ~~~~~~ 
(n=0,1,2,\cdots~ :~a=2^{-1}(2\pi)^{-\frac{1}{2}}L).\label{21'}
\end{equation}
\setcounter{equation}{21}
\renewcommand{\theequation}{\arabic{equation}}
The periodic condition for the static case, Eq.(\ref{21'}) 
indicates that we have other simple poles at $z=((2\pi n)^\frac12 a^{-1})
(\pm 1+i)2^{-\frac{1}{2}}$
(in Eq.(\ref{4'}), $ n=1,2,\cdots,N $
(the finite face value of the periodicity),
i.e., the {\it periodicity dependent} cases) in addition to the triple
pole at $z=0$ (the periodicity independent case) in the integral, 
Eq.(\ref{12}) of the upper half of the complex momentum. Hence we have a different
type of static potential between quark pairs from the periodicity 
independent one, Eq.(\ref{16}).\\
\indent In fact, for the static case, when $c=(2\pi)^{\frac12}$, we have
\begin{eqnarray*}
a^2 z^2 &=& a^2|\vec{k}|^2 \exp(2i\psi)\\ 
         &=& 2n\pi(\cos 2\psi + i\sin2\psi).
\end{eqnarray*}
\noindent If we put $\psi=\frac14 \pi$ or $\psi=\frac34 \pi$, then we get\\
\begin{equation}
\exp(-a^2 z^2)=\exp(-2n\pi(\pm i))=1. \label{22}
\end{equation}
\indent To obtain the static potential between quark pairs, we have to
be careful in calculating the integral I, which  includes simple poles
at $z=((2\pi n)^\frac12 a^{-1})(\pm1 + i)2^{-\frac{1}{2}}$
(the periodicity dependent cases) in addition to the triple pole 
at $z=0$ (the periodicity independent case). Then we have 
\begin{eqnarray}
 I&=&(4\pi)^{-1}\left\{{(-r^{-1}+a^{-2} r)+(-r^{-1})\pi^{-1}
\sum^N_{n=1}n^{-1}\exp(-\kappa_n r)\sin(\kappa_n r)}\right\}\nonumber\\
  &=&(4\pi)^{-1}\left\{-r^{-1}(1+P(a,r,N))+a^{-2} r\right\},\label{23}
\end{eqnarray}
where
$\kappa_n=(\pi n)^\frac12a^{-1}$ and
\begin{equation} 
P(a,r,N)=\pi^{-1}\sum_{n=1}^N n^{-1} \exp(-\kappa_n r)\sin(\kappa_n r).\label{24}
\end{equation}
In Eq.(\ref{23}), $N$ must be finite for the Cauchy's integral formula 
is applied to the calculation of $I$ \cite{11},\cite{12}. This $N$ may be considered
as the finite face value of the periodicity.
And the finiteness of the fundamental constituent at $n \le N<\infty$ as
$$r_n\ge r_N > 0$$
is equivalent to the one of the fundamental constituent at $n \le N<\infty$ as
$$m_n\le m_N < \infty.$$
\noindent The $P(a,r,N)$, Eq.(\ref{24}) is the 
periodicity dependent effect, with $\lim\limits_{a \to \infty}
P(a,r,N)=0$ and $\lim\limits_{r\to0} P(a,r,N)=0$. 
Now, this $P(a,r,N)$ is rewritten as
\begin{equation}
P(a,r,N)=\pi^{-1}\sum_{n=1}^N P_n(a,r,n),\label{25}
\end{equation}
where
\begin{equation}
P_n(a,r,n)=n^{-1}\exp(-\kappa_n r)\sin(\kappa_n r).\label{26}
\end{equation}
Also, $P_n(a,r,n)$, Eq.(\ref{26}) satisfies the following differential equation
\begin{equation}
\left(\frac{d^2}{dr^2}+2n^{\frac12}\frac{d}{dr}+2n\right)
P_n(a,r,n)=0~~~~ (n=1,2,\cdots,N),\label{27}
\end{equation}
where $r$ represents $(\pi^\frac12 a^{-1})r$.\\
\indent Finally, we obtain a static potential between quark 
pairs over all ranges
in the lowest order
\begin{eqnarray}
V(a,r,N)&=&\frac43 g^2_s I\nonumber\\
        &=&\alpha_s \left\{-r^{-1}(1+P(a,r,N))+a^{-2} r \right\}.\label{28}
\end{eqnarray}
\vspace{.5cm}

\noindent{\large\bf 5. Explanation of Top Quark Mass}\\

We are interested in an explanation of the top quark mass
by the periodicity dependent effect in  this model. 
It is to  be noted that $V(a,r,N)$, Eq.(\ref{28}) is  dependent of the 
coupling constant
$\alpha_s$, but $P(a,r,N)$, Eq.(\ref{24}) is irrelevant to 
$\alpha_s$, dominant at short distance and
periodicity dependent. The periodicity dependent 
effect in $V(a,r,N)$ is originated in
$P(a,r,N)$. So our attention is paid to $P(a,r,N)$ in the 
explanation of the top quark mass.\\ 
\indent For the fixed $n$, the periodic condition for the static case,
Eq.(\ref{21'}) gives a relation\\
\setcounter{equation}{20}
\renewcommand{\theequation}{\arabic{equation}$''$}
\begin{equation}
a|\vec k|_n\equiv ak_n = (2\pi)^{\frac12} n^{\frac12}
~~~~(|\vec k|_n\equiv k_n),\label{21''}
\end{equation} 
\setcounter{equation}{28}
\renewcommand{\theequation}{\arabic{equation}}
\hspace{-3mm}and taking the position momentum uncertainty, Eq.(\ref{1}), we obtain
\begin{equation}
a=2(2\pi)^{\frac12} r_n n^{\frac12} = 2(2\pi)^{\frac12} RN^{\frac12}\label{29}
\end{equation}
or
\begin{equation}
L=8\pi r_n n^{\frac12}=8\pi RN^{\frac12},\label{30}
\end{equation}
where $R\equiv r_N$ is the shortest of $r_n$ 
(the closest distance between the quark pair)
and $N$ the largest of $n$ (the finite face value
of the periodicity), respectively. For $L=m_\pi^{-1}$,
from Eq.(\ref{30}) we get 
$$L^{-1}=m_\pi=(8\pi)^{-1}R^{-1}N^{-\frac{1}{2}}$$
and
\begin{equation}
m_t=8\pi m_\pi n^{\frac12}_t~~~(N\equiv n_t, R=m_t^{-1}),\label{31}
\end{equation}
where $N$ may be numerically determined from $P(a,R,N)$ in Eq.(\ref{28}) \cite{13}.\\
\indent The correspondences among $r,n$ and $\kappa_n R$ are as follows:\\
\vspace{-5mm}
\begin{center}
Table 1
\end{center}
\begin{center}
\begin{tabular}{|c|ccccc|}
\hline   
$r$ & $R~(\equiv r_N)$ &$\sim$ & $r_n$ & $\sim$ & $r_1\left(\equiv (8\pi)^{-1}L\right)$\\ \hline
$n$ & $N$            & $\sim$ & $n$   & $\sim$ &  1\\ \hline
~$\kappa_n  R~$ &  ~$(2\cdot2^\frac12)^{-1}~$ &~ $\sim$~ &~ $(2\cdot2^\frac12)^{-1}(\frac{n}{N})^\frac12$~
   & ~$\sim$~&~$(2\cdot2^\frac12)^{-1}(\frac1N)^{\frac12}$~ \\ \hline
\end{tabular}
\end{center}
\noindent From Eqs.(\ref{30}),(\ref{31}) and Table 1, we have the following 
numerical results (including one of the top quark),
for the range of strong interaction $L=1.40~ fm~(=m_\pi ^{-1})$
as shown in Table 2.
\vspace{-3mm}
\begin{center}
Table 2
\end{center}
{ \footnotesize 
\begin{center}
\begin{tabular}{|c|ccc|cccc|p{10in}}\hline
           &   u(d)     & s   &    c   & \multicolumn{2}{c}  b &\multicolumn{2}{c|}  t    \\ \hline
$m_l(\mbox{GeV})$ \cite{14}  &  0.3   &  0.5   &1.35   &   5.0 & (5.08) & (173.43) & (176.23)\\ \hline
$R_l~(fm)$           & 0.6667  & 0.4000  &0.1481&0.0400&
  0.0394 & ~  $1.153\times10^{-3}$ & $1.135\times10^{-3}$ ~\\ \hline
$N_l^\frac12$ & 0.0836 & 0.1393 & 0.3761 & 1.3925 & 1.4142 & 48.31 & 49.09\\ \hline
$~ L(8\pi)^{-1}~(fm) ~$ & ~ 0.0557 & 0.0557 ~  &  0.0557 & 0.0557 & 0.0557  & 0.0557 & 0.0557 \\ \hline
$L~(fm)$ & 1.40 & 1.40 & 1.40 & 1.40 & 1.40 & 1.40 & 1.40 \\ \hline
$N_l$ & 0.0070 & 0.0194 & 0.1415 & 1.9371 & 2 & 2334 & 2410\\ \hline
$P(N_l)$ \cite{13} & 0 & 0 & 0 &  &0.1000 & 0.1850 & 0.1850\\ \hline
\end{tabular} 
\end{center}
}
\vspace{-3mm}
{\footnotesize
$m_l=R^{-1}_l, ~1~\mbox{GeV}^{-1}=0.2fm, ~ m_{\pi}\simeq \frac{1}{7}{\rm GeV},  ~ P(a, R_l, N_l)\equiv P(N_l)$
 and ( ) means a calculated value.
}
\vskip 1em

\noindent In this table, the $P(N_{l})$ is evaluated by the integer value $N_{b}=2$ instead 
of $N_{b0}=1.9371 ~(\sigma N_{b}=( \frac{1.9371}{2}) \cdot 2)$
 and $P(N_{l})$'s for $u(d)$, $s$ and $c$ quarks may be regarded as zero, since
$N_{l}<1$.
In this model, $N_{l}\geq 1$(Eq.(24)).
Table 3 is concerned with the numerically calculated values $P(N)$ at $N$.
 When $P(N_{t})=0.1850 ~(N_{t}=2410)$ may approximately be suitable for the upper
bound of $m_{t}$,
the lower bound of $m_{t}$ may be obtained in such a way that $P(N_{t0})=0.1850 ~ 
(N_{t0}=\sigma N_{t}=2334)$,
where $\sigma$ is defined as  $\sigma=\frac{N_{b0}}{N_{b}}=\frac{N_{t0}}{N_{t}},$ in accordance 
with the $b$ quark case.
For the same periodicity dependent effect on $m_{t}$, between the upper and lower 
bounds of the top quark mass, the $P(N_{t})=P(N_{t0})=0.1850$ have been chosen (see Eq(28)).\\
\vspace{-5mm}
\begin{center}          
Table 3
\end{center}
{\scriptsize
\begin{center}
\begin{tabular}{|c|c|c|c|c|c|c|c|c|c|c|c|}\hline
$N$ & 1 & 2 & 3 & 5 & 10 & 100 & 1000 & 2000 & 2334 & 2410 &$\infty$  \\\hline
$P(N)$ & 0.0784 & 0.1000 & 0.1123  & 0.1263 & 0.1422 & 0.1725 & 0.1833 & 0.1847 & 0.1850 & 0.1850 &0.1884 \\\hline 
\end{tabular}
\begin{eqnarray*}
P(N)   &=& \pi^{-1}\sum\limits_{n=1}\limits^N n^{-1}\exp\left(-\frac{1}{2\cdot2^\frac12}
\left(\frac{n}{N}\right)^\frac12\right)
\sin\frac{1}{2\cdot2^\frac12}\left(\frac{n}{N}\right)^{\frac12},~ RN^\frac12=\frac{L}{8\pi}.
\end{eqnarray*}
\end{center}
}

\noindent From Table 2, it may be found that $b$ and $t$ quarks, belonging to the third
generation of the quark model, have the periodicity dependent effect
between these quark pairs, and that the larger $N_l$ is, the smaller
$\Delta P(N_l)$ is, where $\Delta P(N_l)$ means an increase of $P(N_l)$
in $N_l$. We adopt $P(N_t) = 0.1850$ for the numerically calculated values
$P(N_{t0}) = 0.1850~(N_{t0} = 2334)$ and $P(N_t) = 0.1850~(N_t = 2410)$ as
illustrated in Table 2. Finally, the upper and lower bounds of the top
quark mass have the same order periodicity dependent effect as in Table
2. Thus, from Table 2 we have approximately
\setcounter{equation}{31}
\renewcommand{\theequation}{\arabic{equation}}
\begin{equation}
177~{\mbox{GeV}}> m_t > 173 {\mbox{GeV}}
\end{equation}
for the range of strong interaction $L=1.40fm ~(=m^{-1}_n) $.\\

\vspace{0.5cm}
\noindent {\large \bf 6. Concluding Remarks} \\

In this paper, the phenomenological model for quark pairs (QCD inspired)
beyond the standard model has been developed within the framework of
the quark model with three generations. The closest distance between
the quark pair $r_l$, Eq.(2) is related to the quark mass $m_l =
(r_l^{-1})$ (by definition). This is realized by the constant 
 $L~(=m_{\pi}^{-1})$ and the finite face value of the periodicity $n_l$
as
\setcounter{equation}{29}
\renewcommand{\theequation}{\arabic{equation}$'$}
\begin{equation}
m_l = 8\pi m_{\pi} n_l^{\frac{1}{2}} ~(\mbox{or} ~L=8\pi r_l n_l^{\frac12}),
\end{equation}
where $l$ is one of $n$'s such as $l=u(d),~s,~c,~b$ and $t$, respectively.
In this model, Eq. $(30')$ is valid for $n_l \geq 1$. However, this
equation is still valid for $n_l<1$, numerically. So we may
accept this equation for all $n_l ~(0<n_l<\infty)$, as mentioned above.
The flavor in the quark model with three generations may have such
periodicity in this model as one of the flavor attributes. The 
periodicity independence (or the periodicity dependency) of the
static potential between quark pairs may correspond to the flavor
independence (or the flavor dependency) of it. This may suggest that
there exists a corresponding symmetry or a corresponding broken
symmetry (a source of the heavy quark masses) of it.

The momentum of one quark pair, enclosed by the constant cubic volume
$L^3$, is finite, equal spacing and periodic for periodic boundary
conditions of the quark pair system. The static potential between
such quark pair in the quantum view has been derived, in our fashion,
as follows:

\setcounter{equation}{27}
\renewcommand{\theequation}{\arabic{equation}}
\begin{eqnarray}
V(a,r,N) &=& \alpha_s \{ -{\frac1r} (1+P(a,r,N) )+a^{-2}r \} \\
         & & \left(\alpha_s = {4\over 3} ({g_s^2\over 4\pi}) ,a=2^{-1}
             (2\pi)^{- {1\over 2}} L \right)\nonumber 
\end{eqnarray}
where the periodicity dependent effect $P(a,r,N)$ is

\setcounter{equation}{23}
\begin{equation}
P(a,r,N) = \pi^{-1}\sum_{n=1}^N n^{-1}\exp(-\kappa_nr)
            \sin \kappa_n r 
\end{equation}
     \begin{center}     $(\kappa_n = (\pi n)^{1\over 2} a^{-1})$. \end{center}

\noindent The top quark confined range number (the finite face value of the 
periodicity) $N$ is numerically determined from $P(a,r,N)$ at 
$r=R$, where R is the top quark confined range. Thus from Eq.$(30')$ we
obtain
\setcounter{equation}{30}
\renewcommand{\theequation}{\arabic{equation}}
\begin{equation}
  m_t = 8\pi m_\pi N^{1\over 2},
\end{equation}

\noindent where $m_t = R^{-1}$ and $m_\pi=L^{-1}$, respectively. It is to be
noted that the factor $(8\pi m_\pi )$ in Eq.(31)
(or $r_1^{-1}=\left( {L \over {8 \pi}} \right)^{-1}$ in Table 1) is
another constant in this system.\\
\indent Finally, our numerical results of the top quark mass in Table 2, approximately
\begin{equation}
177~\mbox{\mbox{GeV}}> m_t > 173 ~\mbox{\mbox{GeV}}
\end{equation}
for the range of strong interaction $L=1.40 fm ~(=m_\pi^{-1})$, are consistent 
with the experiments at Fermi Laboratory \cite{16}, \cite{17}.

\vglue .4in
\noindent{\large\bf Acknowledgements}
\vglue .2in

\indent The author would like to thank the particle physics group, Institute 
of Theoretical Physics, SNU and Professors I. T. Cheon, H. W. Lee, and M. Rho (Saclay) 
for their kind discussions. Also, he is much indebted to Professor N. Z. Cho
for numerical calculation.

\vspace{3cm}
\appendix
\setcounter{equation}{0}
\renewcommand{\theequation}{A.\arabic{equation}}

\begin{center}
{\large\bf Appendix}\\
{\bf A derivation of the excitation probability}, $P(-a^2_c k^2)=\exp(a^2_c k^2)$
\end{center}
\indent\indent Consider a particle characterized by a dimensionless 
scalar quantity $-a^2_c k^2\equiv x(0\!<\!x\!<\!\infty)$, associated with 
both the
(finite and equal spacing) squared momentum transfer $-k^2(0<-k^2< \infty)$
in our sense and the finite length parameter $a_c$. Let $P(x)$ 
be a probability that such a particle is excited by $-k^2$ between 
$n'(-k^2)$ and $(n'+1)(-k^2) ~(n'=0,1,2,\cdots)$ without suffering any change.
It is to be noted that the change $\delta(-k^2)$ in $-k^2$ corresponds
to the change $\delta x $ in the probability variable $x$, 
as $-k^2$ does to $x$ (where $\int_0^\infty P(x)dx=1$ 
and $\int_0^\infty P(x)d(-k^2)=a^{-2}_c$ ).
Of course $P(0)=1$, since a   particle has no chance of any changing
in $-k^2\to 0$. On the other hand, $P(x)$ decreases as $-k^2$ increases 
without suffering any change. 
Finally, $P(x)\to 0 $ as $ -k^2\to \infty$.\\
\indent For an infinitesimal dimensionless scalar quantity $\delta x$, 
from the above consideration on $P(x)$, we have
\begin{equation}
P(x+\delta x)\equiv P(x)-P(x)\delta x~~~(\mbox{physically}),\label{a1}
\end{equation}
which is equivalent to
\begin{equation}
P(x)-P(x+\delta x)= P(x)\delta x.\label{a2}
\end{equation}
From (\ref{a2}), we have
\setcounter{equation}{1}
\renewcommand{\theequation}{A.\arabic{equation}$'$}
\begin{equation}
P(x)-[P(x)+\frac{\delta P(x)}{\delta x}\cdot\delta x]= P(x)\delta x.\label{a2'}
\end{equation}
Hence
\setcounter{equation}{2}
\renewcommand{\theequation}{A.\arabic{equation}}
\begin{equation}
\frac{\delta P(x)}{P(x)}=-\delta x \label{a3}
\end{equation}
or
\begin{equation}
P(x)=\gamma\exp(-x). \label{a4}
\end{equation}
\noindent Here the integration constant $\gamma$ can be determined 
by the condition $P(0)=1$. Thus, one obtain $\gamma=1$ and
\begin{equation}
P(-a^2_c k^2)=\exp(a^2_c k^2)~~\left(0<\exp(a^2_c k^2)<1~
\mbox{for}~0<-k^2<\infty\right),\label{a5}
\end{equation}
\noindent where $a_c$ is determined so that our static quark-antiquark potential,
Eq.(\ref{28}) fits experimental data at short or long distances.\\
\vspace{2cm}

\end{document}